\documentclass[10pt, conference, letterpaper]{IEEEtran}

\usepackage{amsfonts}
\usepackage{mathrsfs}
\usepackage{amssymb}
\usepackage{wrapfig}
\usepackage{amsmath}
\usepackage{accents}
\usepackage{dsfont}
\usepackage{amsbsy}
\usepackage{setspace}
\usepackage{graphicx}
\usepackage{subfig}
\usepackage{url}
\usepackage{color}
\usepackage{epstopdf}
\usepackage{amssymb}
\usepackage{framed}
\usepackage{multirow}
\usepackage{mathtools}  
\usepackage{algorithm}
\usepackage{tabularx}
\PassOptionsToPackage{bookmarks={false}}{hyperref}

\IEEEoverridecommandlockouts

\title{Efficient Learning-based Scheduling for Information Freshness in Wireless Networks}
\author{Bin Li\\Department of Electrical, Computer, and Biomedical Engineering\\University of Rhode Island, Kingston, RI 02881, USA\\
Email: binli@uri.edu
\thanks{This research has been supported in part by NSF grants:
CNS-1717108, CNS-1815563, and CNS-1942383.}}
\date{}
\def\argmax{\operatornamewithlimits{arg\,max}}

\begin{document}
\pagestyle{plain}

\maketitle

\newtheorem{theorem}{Theorem}
\newtheorem{lemma}{Lemma}
\newtheorem{claim}{Claim}
\newtheorem{proposition}{Proposition}
\newtheorem{corollary}{Corollary}
\newtheorem{definition}{Definition}
\newtheorem{assumption}{Assumption}
\newtheorem{remarks}{Remarks}

\newcommand{\cS}{\mathcal{S}}
\newcommand{\vS}{\mathbf{S}}
\newcommand{\vQ}{\mathbf{Q}}
\newcommand{\bE}{\mathds{E}}
\newcommand{\mc}{\mathcal}
\newcommand{\mb}{\mathbf}
\newcommand{\bs}{\boldsymbol}
\newcommand{\ol}{\overline}
\newcommand{\wt}{\widetilde}
\newcommand{\wh}{\widehat}
\newcommand{\id}{\mathds{1}}

\makeatletter
\newcommand{\rmnum}[1]{\romannumeral #1}
\newcommand{\Rmnum}[1]{\expandafter\@slowromancap\romannumeral #1@}
\makeatother
\def\maximize{\operatornamewithlimits{Maximize}}
\def\limitsup{\operatornamewithlimits{limsup}}
\def\limitinf{\operatornamewithlimits{liminf}}

\begin{abstract}
Motivated by the recent trend of integrating artificial intelligence into the Internet-of-Things (IoT), we consider the problem of scheduling packets from multiple sensing sources to a central controller over a wireless network. Here, packets from different sensing sources have different values or degrees of importance to the central controller for intelligent decision making. In such a setup, it is critical to provide timely and valuable information for the central controller. In this paper, we develop a parameterized maximum-weight type scheduling policy that combines both the AoI metrics and Upper Confidence Bound (UCB) estimates in its weight measure with parameter $\eta$. Here, UCB estimates balance the tradeoff between exploration and exploitation in learning and are critical for yielding a small cumulative regret. We show that our proposed algorithm yields the running average total age at most by $O(N^2\eta)$. We also prove that our proposed algorithm achieves the cumulative regret over time horizon $T$ at most by $O(NT/\eta+\sqrt{NT\log T})$. This reveals a tradeoff between the cumulative regret and the running average total age: when increasing $\eta$, the cumulative regret becomes smaller, but is at the cost of increasing running average total age. Simulation results are provided to evaluate the efficiency of our proposed algorithm.



\end{abstract}

\section{Introduction}

\begin{figure}[ht!]
	\begin{center}
		\includegraphics[width=0.42\textwidth]{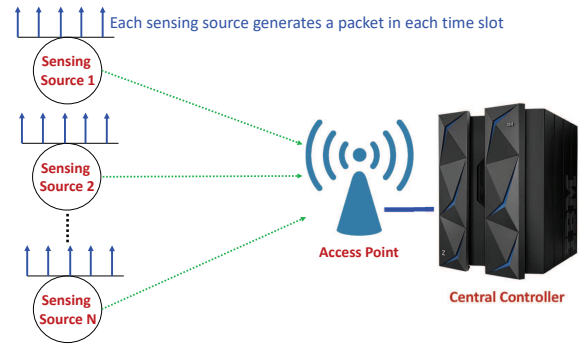}
		\caption{An intelligent Internet-of-Things (IoT) system.}
		\label{fig:SysModel}
	\end{center}
\end{figure}

With the recent advances in artificial intelligence (AI), there is a trend for incorporating AI into the Internet-of-Things (IoT) consisting of multiple wireless sensing sources to provide wise decisions. In such an IoT system, it is critical to make sure that the received sensing information is valuable and timely. As such, in this paper, we consider the problem of scheduling packets from multiple sensing sources to a central controller over a wireless network as shown in Fig. \ref{fig:SysModel}, where packets from different sensing sources have different values or degrees of importance to the central controller for  decision making. 

In particular, we assume that each sensing source constantly generates packets with random values independently and identically distributed (i.i.d.) with an unknown distribution. The value of a packet is revealed only after the central controller successfully receives it. On the one hand, we would like to deliver packets from the most important sensing sources to the central controller for making a better decision subject to the wireless interference constraints. However, the controller does not have any prior knowledge of the degree of importance of these sensing sources and requires to gradually learn these statistics while scheduling the best sensing sources (a.k.a. exploration-exploitation tradeoff in online learning). On the other hand, we should also ensure that the received packets have a low Age-of-Information (AoI) that measures the duration between the packet generation time and its received time. This is because the stale information is less useful to the central controller and might even mislead the controller to make harmful decisions. To that end, we aim to develop a scheduling algorithm to achieve this dual objective, which is complicated by the strong coupling between the learning and AoI performance. 

Without the AoI constraint, the considered problem can be formulated as a combinatorial multi-armed bandit (MAB) problem (e.g., \cite{anantharam1987asymptotically,gai2012combinatorial,chen2013combinatorial,combes2015combinatorial}), where each arm corresponds to a sensing source, and the goal is to minimize the cumulative regret over a finite time horizon (i.e., the difference between the optimal cumulative reward and the cumulative reward under an algorithm). The combinatorial MAB algorithms allow to play multiple arms simultaneously in each time slot instead of one arm as in classical MAB algorithms such as Upper-Confidence-Bound (UCB \cite{auer2002finite}), Kullback-Leibler UCB (KL-UCB \cite{garivier2011kl}), and Thompson sampling \cite{agrawal2012analysis}. The efficient combinatorial MAB algorithms should quickly identify the set of best arms and keep pulling them. This, however, leads to the large AoI for other relatively poor arms. In particular, the AoI keeps increasing over time and this implies that the received information from relatively poor arms is outdated. This motivates us to incorporate the AoI metric into the learning algorithm design for combinatorial MAB problems. 

While there are some recent works on AoI-efficient wireless scheduling (e.g, \cite{lu2018age,kadota2018optimizing,kadota2019minimizing} and see \cite{sun2019age} for an overview), their goal was to minimize AoI while guaranteeing the desired throughput. They did not consider the MAB setting with unknown system statistics, which is typical in intelligent IoT systems. As such, in this paper, we integrate the main ideas of UCB algorithm (e.g., \cite{auer2002finite}) and AoI-efficient scheduling (e.g., \cite{lu2018age}), and propose a Learning-based Age-Efficient Scheduling (LAES) Algorithm that utilizes both the UCB estimates and AoI metrics. While there are some recent works in combinatorial bandits with fairness constraints (e.g., \cite{li2019combinatorial}), they focused on the long-term fairness constraint, i.e., each arm should at least be played for a fixed fraction of times on average. The main approach is to maintain a virtual queue for each arm that keeps track of its debt and prioritizes arms with high virtual queue lengths, typically referred to as virtual queue techniques (e.g., see \cite{neely2010stochastic} for an overview). However, the AoI captures the short-term dynamic of the system and thus its evolution is fundamentally different from the virtual queue length. In particular, it has an unbounded decrement whenever a packet is successfully delivered, which has a significant impact on the performance analysis of the proposed algorithm. 
Our contributions in this work are summarized as follows:

$\bullet$ We develop a parameterized maximum-weight type scheduling policy that combines both the AoI metric and UCB estimate in its weight measure (cf. Section \ref{sec:algorithm}). In particular, we use the parameter $\eta$ to balance the AoI metric and UCB estimate. The larger the $\eta$, the more emphasis on the UCB estimate and thus leads to the smaller regret, but it is at the cost of the larger AoI.

$\bullet$ We derive an upper bound on the running average total age under our proposed algorithm with any $\eta>0$ (cf. Proposition \ref{prop:age}), which linearly increases with the parameter $\eta$. Such an upper bound is tight in some cases in the sense that the average total average under our proposed algorithm linearly scales with the parameter $\eta$.


$\bullet$ We show that the cumulative regret over a finite time horizon $T$ can be bounded from above by $O(NT/\eta+\sqrt{NT\log T})$\footnote{$f(x)=O(x)$ if there exists a positive real number $M$ such that $f(x)\leq Mx, \forall x\geq0$.} under our proposed algorithm (cf. Proposition \ref{prop:regret}). Here, the second term has the same order as that of the UCB algorithm and is attributed to the cost for exploration/exploitation in online learning, while the first term $NT/\eta$ is the cost paid for improving AoI performance. This, together with the derived upper bound on the running average total age, reveals a tradeoff: when  increasing $\eta$,  the  regret  upper bound  decreases,  but  the  upper  bound  on  running average  total  age increases. 

$\bullet$ We support our analytical results with extensive simulations (cf. Section \ref{sec:simulation}), which demonstrates the superior performance of our proposed algorithm over both UCB algorithm and age-based algorithm (i.e., our proposed algorithm with $\eta=0$). Simulation results also confirm a tradeoff between the cumulative regret and the running average total age. The desired tradeoff can be achieved by tuning the value of parameter $\eta$.

The remainder of this paper is organized as follows: Section \ref{sec:relatedwork} reviews related work. Section \ref{sec:model} introduces system model and problem statement. Section \ref{sec:algorithm} introduces our proposed algorithm, and analyzes both its AoI and regret performance. Section \ref{sec:simulation} presents simulation results and Section \ref{sec:conclusion} concludes this paper.

\section{Related Work and Context}
\label{sec:relatedwork}
In this section, we overview two main areas that are closely related to our work: multi-armed bandit and age of information, and further provide a brief discussion of our design methodology in the context of prior work.

\textbf{(a) Multi-Armed Bandit:} The MAB problem models an agent that attempts to learn system statistics while optimizing its decision based on existing learning experiences, and has wide applications in recommender systems, healthcare, finance, and computer networks. As such, it has been received extensive research efforts (e.g., \cite{lai1985asymptotically,auer2002finite,garivier2011kl,agrawal2012analysis}). The seminal work of Lai and Robbins \cite{lai1985asymptotically} established a fundamental logarithmic lower bound on the cumulative regret (i.e., the difference between the optimal cumulative reward and the cumulative reward under an algorithm) over a finite time horizon under a class of uniformly good policies and developed a UCB algorithm that asymptotically achieves this fundamental lower bound. Such a logarithmic regret bound has been shown to be achieved by the sample-mean-based UCB algorithm and $\epsilon$-greedy policy (see \cite{auer2002finite}), Kullback-Leibler UCB (KL-UCB \cite{garivier2011kl}), and Thompson sampling \cite{agrawal2012analysis}.

Subsequent works extended the classical MAB problem to various settings that account for different applications. The one closest to ours is combinatorial MAB (e.g., \cite{anantharam1987asymptotically,gai2012combinatorial,chen2013combinatorial,combes2015combinatorial}), where a subset of arms can be played simultaneously at each time. More recent works considered the combinatorial MAB with fairness constraint (e.g., \cite{li2019combinatorial,patil2020achieving}), where each arm should at least be played for a certain fraction of time on average. The authors introduced the  virtual-queue-length to address fairness constraint and incorporated it into the algorithm design. However, all these MAB works did not address the AoI performance and thus yet unbounded AoI over time, as demonstrated in our simulations (cf. Section \ref{sec:simulation}).

\textbf{(b) Age of Information:} AoI measures the duration between the time when the information was generated and its received time. It directly captures the information freshness and thus has received great attention in recent years. Unlike the traditional queueing delay that is negligible in the case with a low sampling rate (i.e., low arrival rate), the AoI is dominated by the inter-arrival time and thus is rather large in the low sampling rate regime. This key difference has spurred AoI research in several aspects in recent years, e.g., AoI analysis and optimization (e.g., \cite{kaul2012real,altman2019forever}), AoI in vehicular networks (e.g., \cite{kaul2011minimizing,choudhury2020experimental}), online sampling and remote estimation (e.g., \cite{nar2014sampling,ornee2019sampling}), AoI and energy harvesting (e.g., \cite{yates2015lazy,sun2017update,dong2020energy}), just to name a few. 

The one that is closest to our research is the AoI-efficient scheduling in wireless networks (e.g., \cite{lu2018age,kadota2018optimizing,kadota2019minimizing} and see \cite{sun2019age} for an overview) that aims to develop wireless scheduling algorithms with the goal of minimizing AoI. For example, the authors in \cite{lu2018age} developed an age-based scheduler for real-time traffic that achieves not only desired timely throughput but also guaranteed AoI performance. Our research differs from this line of research in that we explicitly incorporate AoI metrics into the MAB algorithm design, which is desirable in the emerging intelligent IoT applications. This key difference poses significant challenges in guaranteeing information freshness in the MAB setting that is unseen in existing AoI research. While a recent work \cite{fatale2020regret} considered the AoI performance in the MAB setting, it focused on the single-user setting and did not consider the case with multiple users and wireless interference constraints. 

\textbf{(c) Our Design Philosophy:} In this paper, we extend a UCB-type algorithm to our setting that demands desired AoI performance while minimizing cumulative regret over time. One extreme is to serve arms with the largest UCB estimates in order to minimize the cumulative regret, but it can result in increasing AoI over time. The other extreme is to serve arms with the largest ages, yet this could lead to a large regret. This is because it does not learn any system statistics nor exploit the best arms so far. Therefore, it is clear that one should tradeoff the benefits of these two approaches. The natural idea is to integrate both UCB estimates and AoI metrics into the scheduling decisions. However, the AoI metric in our work is fundamentally different from the virtual queue length, since it has an unbounded decrement whenever a packet is successfully delivered. Such an abrupt dynamic poses a significant challenge in characterizing AoI performance. The main contribution of this paper is to develop a parameterized learning-based age-efficient algorithm and to show that such an algorithm achieves a tradeoff between the cumulative regret and average total age, which can be tuned by our algorithmic parameter.


\section{system model}
\label{sec:model}
We consider a wireless network with $N$ links, where each link represents a transmitter-receiver pair that are within the transmission range of each other. We assume that the system operates in slotted time with normalized slots $t\in\{0,1,2,\ldots\}$. In each time slot $t$, the transmitter of link $n$ ($n=1,2,\ldots,N$) generates a packet with a random value $X_n(t)\in[0,1]$, which is independently and identically distributed (i.i.d.) with an unknown mean $\mu_n$. Here, $X_n(t)$ 
represents the \emph{reward} when a packet is successfully delivered over link $n$ in time slot $t$. Due to the wireless interference constraints, only a subset of links can transmit in each time slot. We use $S_n(t)=1$ if link $n$ is scheduled for transmission in time slot $t$, and $S_n(t)=0$ otherwise. We call $\mb{S}(t)\triangleq(S_n(t))_{n=1}^{N}$ the \emph{feasible schedule} denoting the set of links that can be active simultaneously in time slot $t$. Let $\mc{S}$ be the collection of all feasible schedules. We assume that each link $n$ experiences i.i.d. ON-OFF channel fading over time with $C_n(t)=1$ denoting that the channel of link $n$ is ON in time slot $t$. Let $p_n\triangleq\Pr\{C_n(t)=1\}$ be the probability that link $n$ has an available channel in time slot $t$. We assume that each link has a non-zero probability that its channel is ON, i.e., $p_{\min}\triangleq\min_{n}p_n>0$. Hence, the received reward $R(t)$ in each time slot $t$ can be expressed as $R(t)\triangleq\sum_{n=1}^{N}X_n(t)C_n(t)S_n(t)$. We consider the case where the channel state is known via channel probing at the beginning of each time slot\footnote{Our algorithm design and its analysis can be easily adapted to the case with unknown channel state.}.

Our goal is to maximize the cumulative reward $\sum_{t=0}^{T-1}R(t)$ until the $T^{th}$ time slot while guaranteeing the desired information freshness. If the statistics of rewards (i.e., $\{\mu_n, n=1,2,\ldots,N\}$) are known in advance, then the first objective can be achieved by solving the following optimization problem:
\begin{align}
\mb{S}^{*}(t)\triangleq(S_n^*(t))_{n=1}^{N}\in\argmax_{\mb{S}\in\mc{S}} \sum_{n=1}^{N}\mu_n C_n(t)S_n. 
\end{align}
That is, it serves a set of non-interfering and available links with the maximum sum of mean rewards in each time slot. Unfortunately, the statistics of rewards are unknown. This requires the algorithm not only to learn these statistics (also known as (a.k.a.) exploration) but also to select the best schedule so far (a.k.a. exploitation). Our first goal is equivalent to minimizing the \emph{cumulative regret} over consecutive $T$ time slots, which is the gap between the accumulated reward and the optimal reward, i.e.,
\begin{align*}
\text{Reg}(T) \triangleq \sum_{t=0}^{T-1}\sum_{n=1}^{N}\left(\bE\left[\mu_nC_n(t)S^*_n(t)\right]-\bE\left[\mu_nC_n(t)S_n(t)\right]\right).
\end{align*}

\begin{figure}[ht!]
	\begin{center}
		\includegraphics[width=0.42\textwidth]{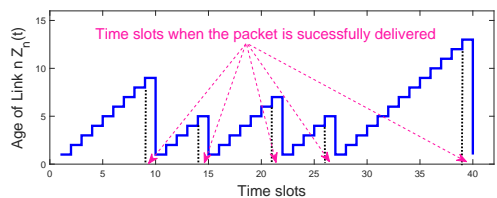}
		\caption{The evolution of age of link $n$.}
		\label{fig:AgeDynamic}
	\end{center}
\end{figure}

To address our second goal for the desired information freshness, we introduce $Z_n(t)$ to denote the \emph{age} of information received from the $n^{th}$ link in time slot $t$, which increases by one if a packet is not received by the receiver of link $n$ in time slot $t$ and reset to one otherwise, i.e., 
\begin{align}
\label{eqn:age:dynamics}
Z_n(t+1) = 
\begin{cases}
Z_n(t) + 1  & \text{if $S_n(t)C_n(t)=0$}; \\
1           & \text{if $S_n(t)C_n(t)=1$}.
\end{cases}
\end{align}
Fig. \ref{fig:AgeDynamic} shows one sample path of age of link $n$. We can obverse from Fig. \ref{fig:AgeDynamic} that $Z_n(t)$ resets to one whenever there is a successful packet delivery. We note that the dynamics of the age is similar to that of Time-Since-Last-Service (TSLS) counter in \cite{li2013throughput,li2014throughput}.

Our second goal is to keep information as fresh as possible, i.e., minimizing $\sum_{n=1}^{N}\bE[Z_n(t)]$. We achieve this dual objective by developing a parametric class of wireless schedulers that efficiently utilize a combination of UCB estimates for minimizing the cumulative regret and ages in its decision. 


\section{Algorithm Design and Performance Analysis}
\label{sec:algorithm}
In this section, we develop a learning-based wireless scheduler by integrating the key idea of the well-known UCB algorithms (see \cite{auer2002finite}) and age metrics. In particular, the UCB is utilized to deal with the fundamental exploitation-exploration tradeoff in online learning and aims to achieve minimum cumulative regret. On the other hand, age metrics are employed to guarantee desired information freshness.

In order to obtain the weight for exploitation and exploration, we introduce the following notations. Let $H_n(t)$ be the number of times link $n$ has successfully received a packet until time slot $t$, i.e., $H_n(t)\triangleq\sum_{\tau=0}^{t-1}C_n(\tau)S_n(\tau)$. We set $H_n(0)=0$ due to the fact that the system starts at $t=0$. We use $\ol{\mu}_n(t)$ to denote the sample mean of the received rewards of link $n$ until time slot $t$, i.e., $\ol{\mu}_n(t)\triangleq\left(\sum_{\tau=0}^{t-1}X_n(\tau)C_n(\tau)S_n(\tau)\right)/H_n(t)$. If $H_n(t)=0$ (i.e., link $n$ has not successfully received a packet yet until time slot $t$), we set $\ol{\mu}_n(t)=1$. Let $w_n(t)$ denote the \emph{UCB estimate} of link $n$ in time slot $t$ and is defined as follows:
\begin{align}
w_n(t)\triangleq \min\left\{\ol{\mu}_n(t)+\sqrt{\frac{3\log t}{2H_n(t)}}, 1\right\},
\end{align}
where $\sqrt{3\log t/(2H_n(t))}$ is the exploration term that measures the uncertainty of the received reward of link $n$ until time slot $t$. Indeed, the smaller the $H_n(t)$, the less exploitation of link $n$ and thus less accuracy of its sample mean estimation, in which case link $n$ should get a higher priority to be scheduled. Here, we use the truncated version of the UCB estimate, since the actual reward of each link is at most $1$. Again, when $H_n(t)=0$, we set $w_n(t)=1$. That is, if link $n$ has not been scheduled yet until time slot $t$, it has the highest priority to get served. 

In order to achieve a low cumulative regret, we prefer to serve links with large UCB estimates in each time slot. Indeed, we would like to serve links with high sample mean rewards and links with large uncertainties of received rewards due to fewer explorations. In order to address information freshness guarantees, we also need to incorporate age metrics into the scheduling design. In particular, the links with large ages should get high priorities to be scheduled. This naturally motivates the following algorithm. 

\begin{algorithm}
\caption{Learning-based Age-Efficient Scheduling (LAES) Algorithm}
In each time slot $t$, given the channel state $(C_n(t))_{n=1}^{N}$, select a schedule $\wh{\mb{S}}[t]\triangleq(\wh{S}_n(t))_{n=1}^{N}$ satisfying
\begin{align}
\wh{\mb{S}}[t]\in\argmax_{\mb{S}\in\mc{S}}\sum_{n=1}^{N}\left(Z_n(t)+\eta w_n(t)\right)C_n(t)S_n,
\end{align}
where $\eta\geq0$ is some control parameter. 
\end{algorithm}

Note that $\eta$ is a parameter that balances the age metrics and the UCB estimates. In particular, if $\eta=0$, then the LAES coincides with the age-based policy (see \cite[Ch. 4.5.4]{sun2019age}). In the presence of fully-connected networks with non-fading channels (i.e., at most one link can be scheduled in each time slot and $C_n(t)=1, \forall n, t\geq0$), the age-based policy is equivalent to the well-known Round-Robin policy that serves links in turn. In such a case, the age-based policy, in fact, minimizes the average total age (see \cite[Proposition 2]{li2013throughput}). The larger the $\eta$, the higher priority the UCB estimates and hence yields a smaller cumulative regret. 

Next, we characterize the age performance of the proposed LAES Algorithm. 


\begin{proposition} 
\label{prop:age}
[Information Freshness Guarantee] If $Z_n(0)=0, \forall n=1,2,\ldots,N$, then, under the LAES algorithm with any $\eta\geq0$, the running average total age can be bounded from above as follow:
\begin{align*}
\frac{1}{T}\sum_{t=0}^{T-1}\sum_{n=1}^{N}\bE\left[Z_n(t)\right]\leq \frac{(\eta+1)N^2}{p_{\min}},
\end{align*}
holding for any $T\geq1$, where $p_{\min}\triangleq\min_{n}p_n>0$.
\end{proposition}
\begin{IEEEproof} We consider the Lyapunov function $V(t)$
\begin{align}
V(t) \triangleq \sum_{n=1}^{N}Z_n(t).
\end{align}
and study its drift. Using telescoping techniques as in the classical Lyapunov drift analysis (e.g., \cite{neely2010stochastic}), we obtain an upper bound on the running average total age. Please see Appendix \ref{App:proof:age} for the detailed proof.
\end{IEEEproof}

\begin{remarks}
From Proposition \ref{prop:age}, we can see that the running average total age is bounded under the LAES Algorithm with any $\eta\geq0$, which is desirable since the central controller always demands a certain degree of information freshness. In addition, the derived upper bound on the running average total age linearly increases with the parameter $\eta$. This matches our intuition on the LAES Algorithm that a large $\eta$ implies a smaller weight on the age metric and thus deteriorates the AoI performance. 
\end{remarks}
\begin{remarks}
The derived upper bound on the running average total age linearly scales with the parameter $\eta$, which might be tight in some cases. Indeed, consider two interfering non-fading links, where $C_n(t)=1, \forall n=1,2, \forall t$, and at most one link can be scheduled in each time slot. Suppose $\mu_1 > \mu_2$, and  assume that both links are scheduled sufficiently many times. In such a case, both $w_1(t)$ and $w_2(t)$ are close to $\mu_1$ and $\mu_2$, respectively. As such, under the LAES Algorithm, link 2 is scheduled roughly one every $\lceil\eta(\mu_1-\mu_2)\rceil$ time slots and link 1 is scheduled in all other time slots. Hence, the average age of link 2 in each time slot is roughly equal to $(1+2+3+\ldots+\lceil\eta(\mu_1-\mu_2)\rceil)/\lceil\eta(\mu_1-\mu_2)\rceil=(1+\lceil\eta(\mu_1-\mu_2)\rceil)/2$. On the other hand, the average age of link 2 in each time slot is equal to $(\lceil\eta(\mu_1-\mu_2)\rceil-1 + 2)/\lceil\eta(\mu_1-\mu_2)\rceil=1+1/\lceil\eta(\mu_1-\mu_2)\rceil$. Hence, the average total age in each time slot is $O(\eta)$. 
\end{remarks}

\begin{remarks}
In the case that all links have a non-zero probability of the channel being OFF (i.e., $p_n<1, \forall n=1,2,\ldots,N$), the upper bound on the mean total age is independent of the parameter $\eta$. Indeed, if the event $\mc{F}_n(\tau)\triangleq\{C_n(\tau)=1, C_{n'}(\tau)=0, \forall n'\neq n\}$ happens for some $\tau\in[t-m+1,t)$, then under the LAES Algorithm, link $n$ should be scheduled at least once during the past $m$ time slots, and thus $Z_{n}(t)<m$. This implies that 
\begin{align}
&\Pr\left\{Z_n(t)\geq m\right\}\nonumber\\
\leq&\Pr\left\{\mc{F}_n(\tau) \text{ does not happen for all }\tau\in[t-m+1,t)\right\}\nonumber\\
\stackrel{(a)}{=}&\nu_n^m\stackrel{(b)}{\leq} \nu^m,
\end{align}
where step $(a)$ is true for $\nu_n\triangleq 1- p_n\Pi_{n'\neq n}(1-p_{n'})\in (0,1)$ under the assumption that $p_{n}<1, \forall n$, and follows from the fact that channel rates are independently distributed across links and i.i.d. over time for each link, and $(b)$ holds for $\nu\triangleq \max_{n}\nu_n$. Hence, we have 
\begin{align}
\label{eqn:age:fading:ub}
\bE\left[Z_n(t)\right]=\sum_{m=1}^{\infty}\Pr\{Z_n(t)\geq m\}\leq\sum_{m=1}^{\infty}\nu^m=\frac{\nu}{1-\nu}.
\end{align}
As such, the average total age in each time slot is upper bounded by $N\nu/(1-\nu)$, which is independent of parameter $\eta$. However, such an upper bound is extremely large, as demonstrated in Section \ref{sec:simulation}, and thus it does not say too much on the dependence of the average total age on the parameter $\eta$ when the average age is moderate or small. 
\end{remarks}


\begin{figure*}[!htbp]
\centering \subfloat[Cumulative Regret]{
\label{fig:sim:nonfading:regret}
\includegraphics[scale=0.4]{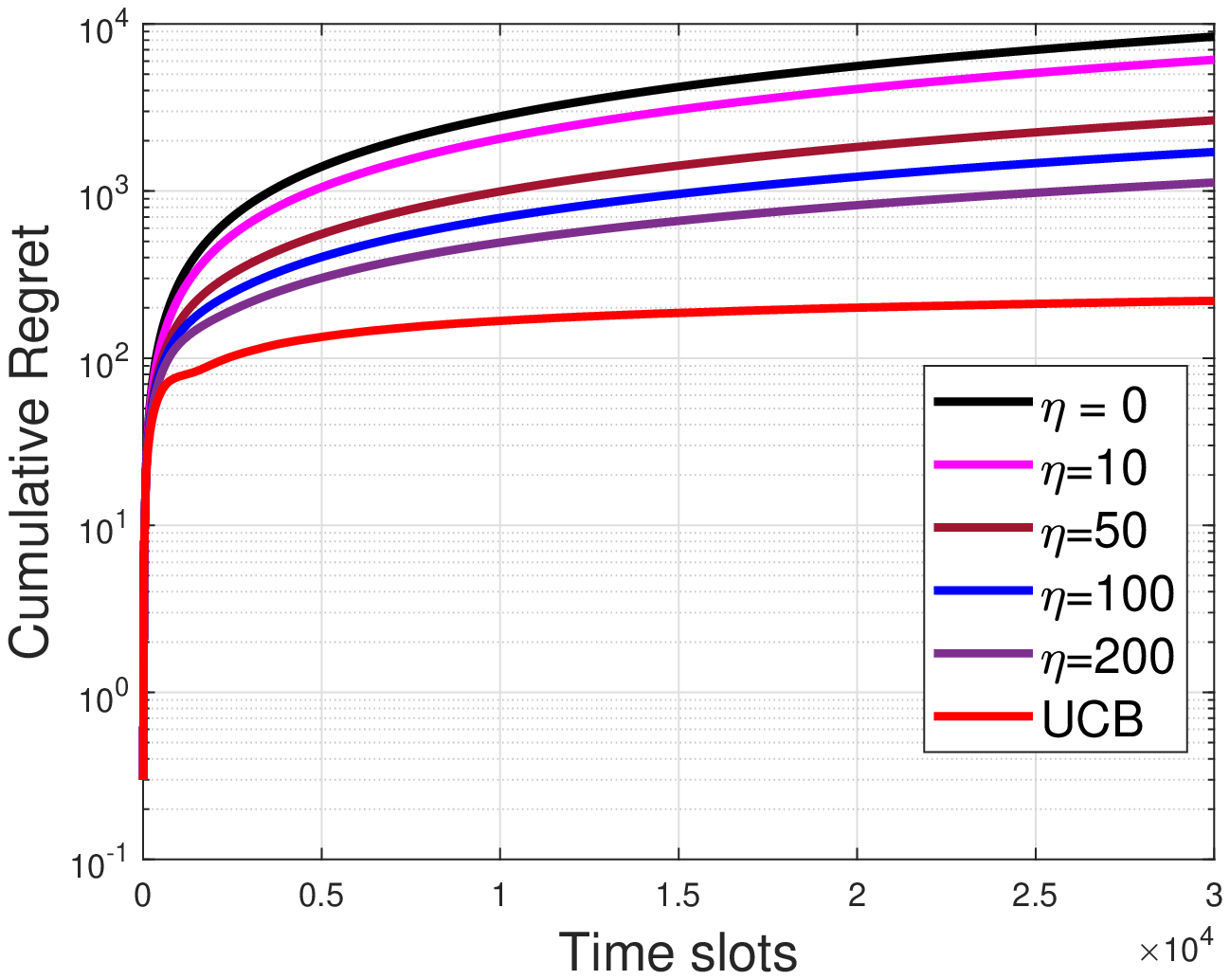}
\hspace{-0.2in}
} \subfloat[Successful Delivery Ratio]{ \label{fig:sim:nonfading:ratio}
\includegraphics[scale=0.4]{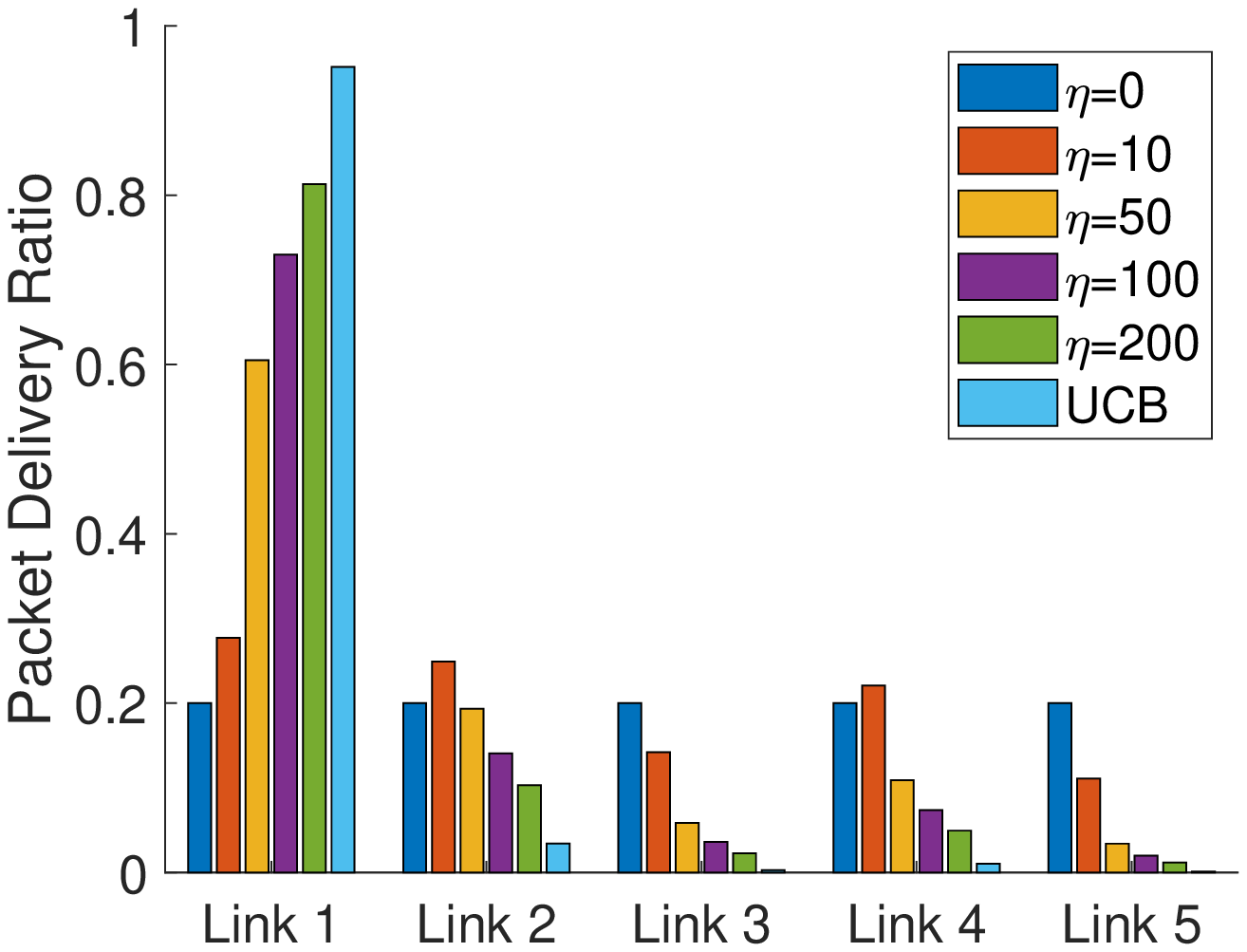}
\hspace{-0.2in}
} 
\subfloat[Total Average Age]{
\label{fig:sim:nonfading:age}
\includegraphics[scale=0.4]{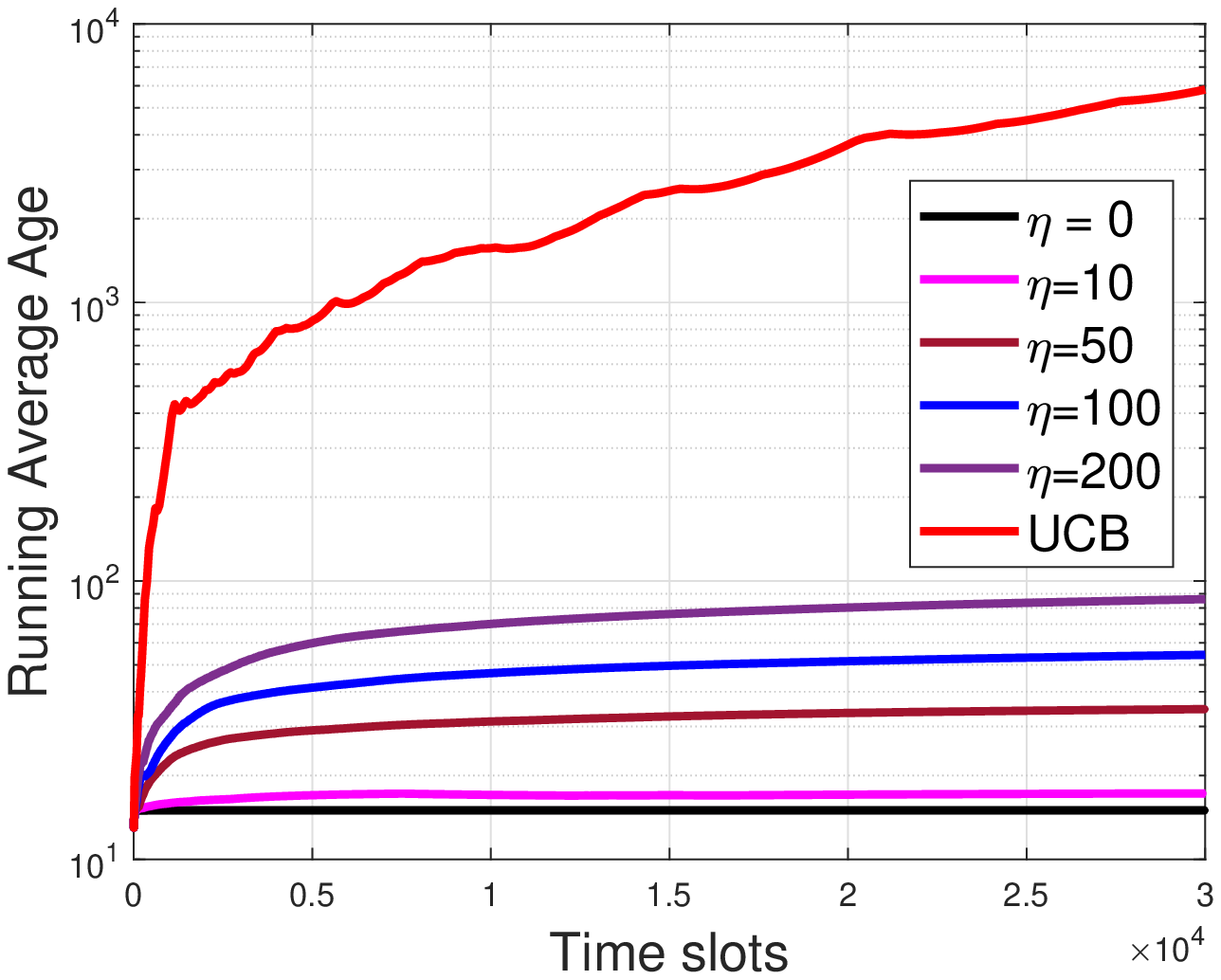}
} 
\caption{Performance of the LAES Algorithm in a fully-connected non-fading network.}
\label{fig:sim:nonfading}
\end{figure*}

\begin{figure*}[!htbp]
\centering \subfloat[Cumulative Regret]{
\label{fig:sim:fading:regret}
\includegraphics[scale=0.4]{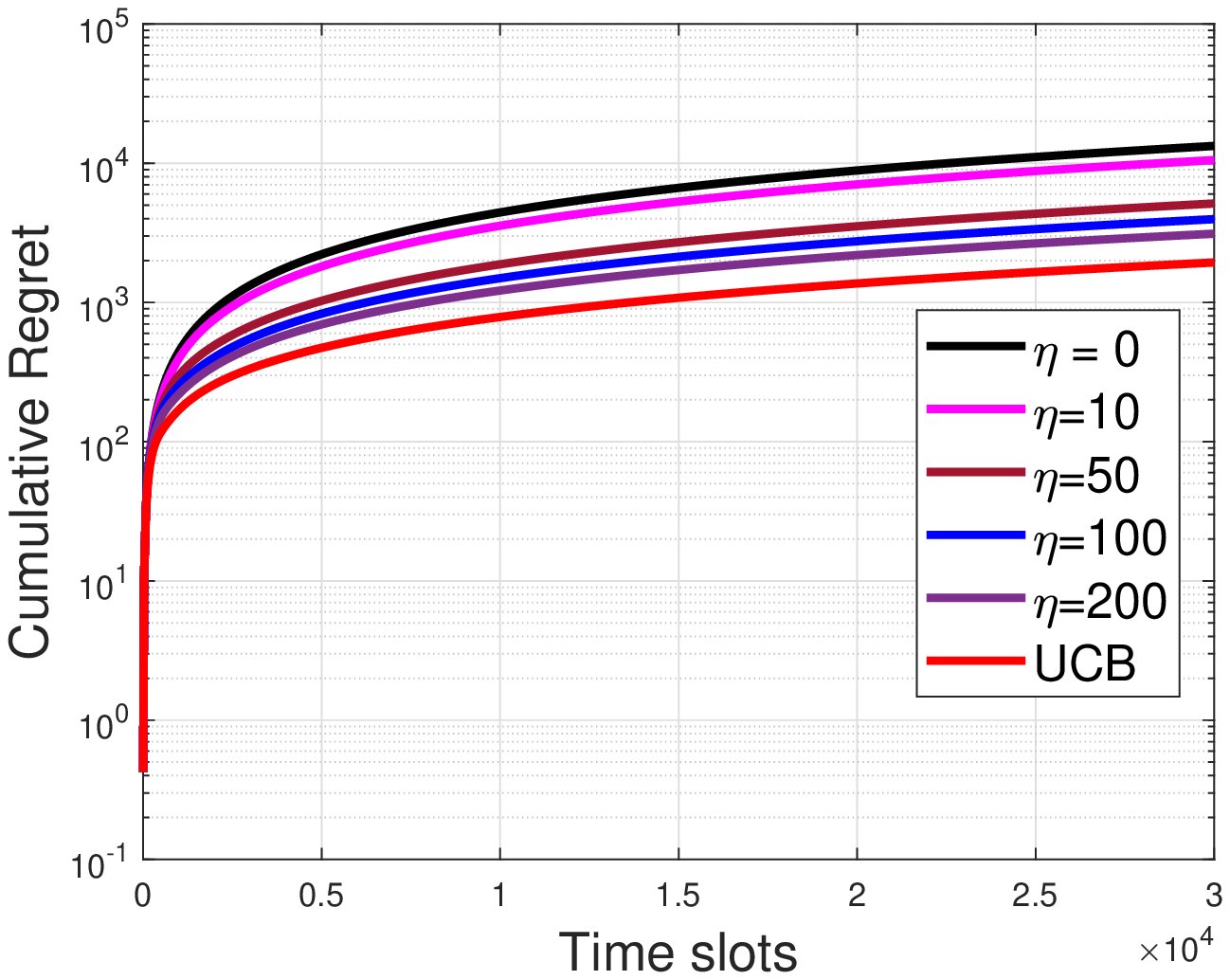}
\hspace{-0.2in}
} \subfloat[Successful Delivery Ratio]{ \label{fig:sim:fading:ratio}
\includegraphics[scale=0.4]{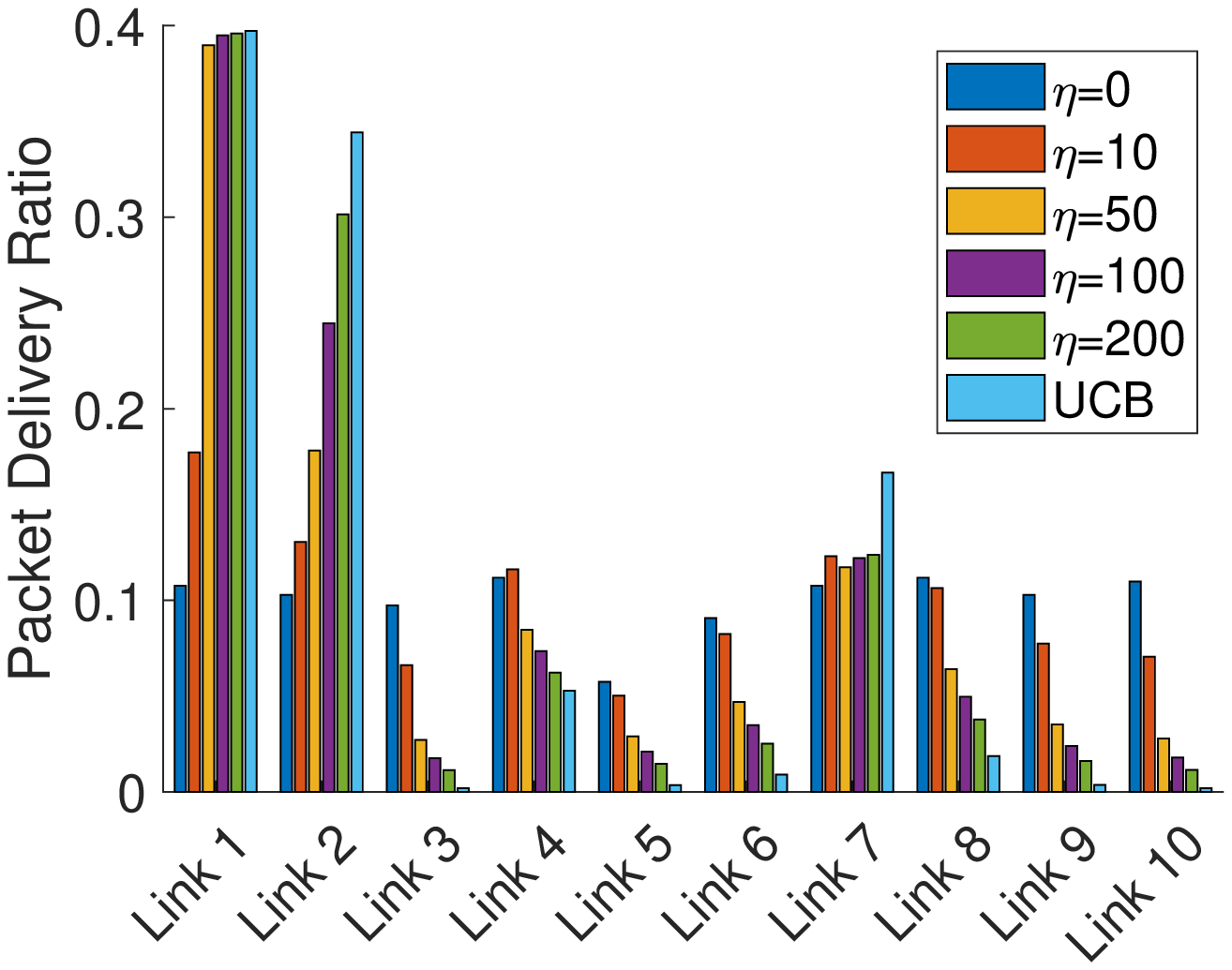}
\hspace{-0.2in}
} 
\subfloat[Total Average Age]{
\label{fig:sim:fading:age}
\includegraphics[scale=0.4]{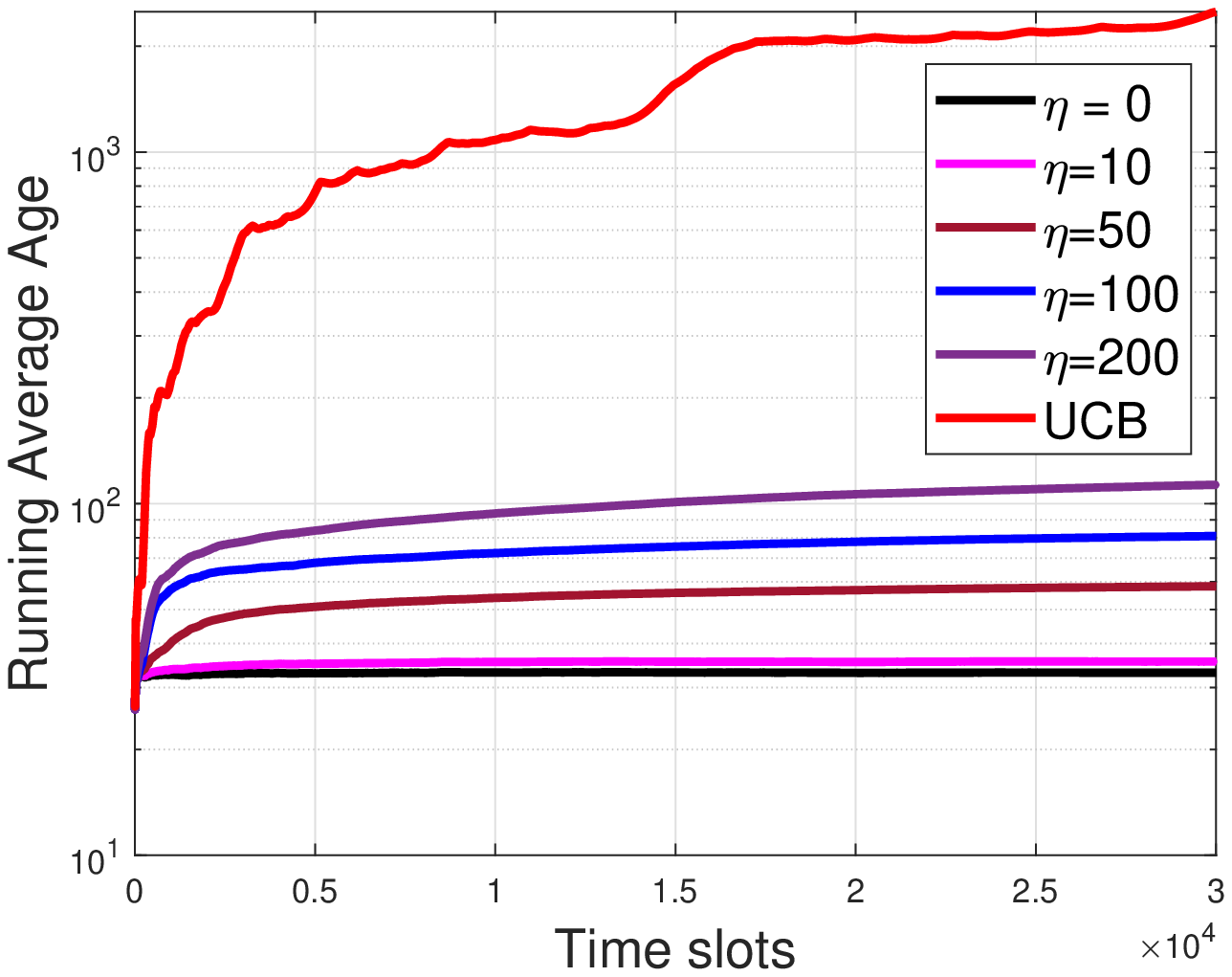}
} 
\caption{Performance of the LAES Algorithm in a $10-$link ON-OFF fading network.}
\label{fig:sim:fading}
\end{figure*}

Lastly, we provide an upper bound on the cumulative regret under the LAES Algorithm with $\eta>0$.

\begin{proposition}
\label{prop:regret}
[Upper Bound on Regret] If $Z_n(0)=0, \forall n=1,2,\ldots,N$, then, under the LAES Algorithm with $\eta>0$, the cumulative regret $\text{Reg}(T)$ until time slot $T>0$ can be bounded from above as follows:
\begin{align*}
R(T)\leq\frac{NT}{\eta}+2\sqrt{6N|\mb{S}|_{\max}T\log T}+N\left(1+\frac{5\pi^2}{12}\right),
\end{align*}
where $|\mb{S}|_{\max}$ denotes the maximum number of links that can be scheduled simultaneously in each time slot.
\end{proposition}
\begin{IEEEproof}
We first perform drift-plus-penalty analysis on 
\begin{align}
\bE\left[V(t+1)-V(t)\right]+\eta\Delta R(t),
\end{align}
where $\Delta R(t)\triangleq\sum_{n=1}^{N}\bE\left[\mu_nC_n(t)S_n^*(t)-\mu_nC_n(t)\wh{S}_n(t)\right]$ and the cumulative regret $\text{Reg}(T)\triangleq\sum_{t=0}^{T-1}\Delta R(t)$. Then, we carefully incorporate the regret analysis for classical UCB algorithm (e.g., \cite{auer2002finite}) into our analysis. The analysis is similar to the line of regret analysis in \cite{hsu2018integrate} and \cite{li2019combinatorial}, and is available in Appendix \ref{App:proof:regret}.
\end{IEEEproof}

\begin{remarks}
The derived upper bound on the cumulative regret consists of two terms: (i) $2\sqrt{6N|\mb{S}|_{\max}T\log T}+N(1+5\pi^2/12)$ has the same order $O(\sqrt{NT\log T})$ as the instance-independent upper bound for the classical UCB algorithm (see \cite[Ch. 2.4.3]{bubeck2012regret}) and thus this term is attributed to the cost involved in the exploration/exploitation process in online learning; (ii) $NT/\eta$ decreases as parameter $\eta$ increases. This also matches our intuition on the LAES Algorithm: the larger the $\eta$, the larger the weight put on the UCB estimate and thus yields a smaller cumulative regret. 

This together with Proposition \ref{prop:age} reveals a tradeoff between the running average total age and the regret performance in the general network setup under the LAES Algorithm: when increasing $\eta$, the regret upper bound decreases, but the upper bound on running average total age increases. That is, the improvement of the cumulative regret is at the cost of increasing running average total age. Moreover, it can be easily derived that the product of the upper bound on the running average total age and the upper bound on the cumulative regret is on the order of $O(N^3T)$ for any $\eta\leq O(\sqrt{NT/\log T})$. In Table \ref{table:tradeoff}, we provide three different $\eta$ values to illustrate the tradeoff between the running average total age and cumulative regret.

\begin{table}[h!]
\centering
\begin{tabular}{ | c | c| c | } 
\hline
Parameter $\eta$ & Regret  & Age \\ 
\hline
\hline
$O(\sqrt{NT/\log T})$ & $O(\sqrt{NT\log T})$ & $O(\sqrt{N^5T/\log T})$ \\ 
\hline
$O(\sqrt[3]{NT})$ & $O(\sqrt[3]{N^2T^2})$ & $O(\sqrt[3]{N^7T})$ \\ 
\hline
$O(1)$ & $O(NT)$ & $O(N^2)$ \\ 
\hline
\end{tabular}
\caption{Cumulative Regret vs. Running Average Age.}
\label{table:tradeoff}
\end{table}
From Table \ref{table:tradeoff}, we can see that in order for the cumulative regret to be on the same order as that for UCB algorithm (i.e., $\text{Reg}(T)=O(\sqrt{NT\log T})$), the running average total age should be on the order of $O(\sqrt{N^5T/\log T})$ under the LAES Algorithm. Nevertheless, we can use a relatively large $\eta$ (e.g., $\eta=200$) and achieve both low regret and low average age, as demonstrated in Section \ref{sec:simulation}. 
\end{remarks}

\section{Simulations}
\label{sec:simulation}

In this section, we perform simulations to evaluate the performance of our proposed LAES Algorithm. We consider the following two network setups: (i) a fully-connected non-fading network with $N=5$ links (at most one link can be scheduled in each time slot and $C_n(t)=1,\forall n, t\geq0$), and (ii) a $10-$link ON-OFF fading network where at most two links can be scheduled in each time slot. For the first setup, the mean reward vector is $\bs{\mu}=(0.9, 0.8, 0.5, 0.7, 0.2)$. For the second network setup, we set the mean reward vector $\bs{\mu}=(0.9, 0.8, 0.4, 0.7, 0.5, 0.6, 0.75, 0.65, 0.5, 0.4)$ and heterogeneous ON-OFF channel fading parameters $\mb{p}=(0.8, 0.7, 0.6, 0.9, 0.2, 0.5, 0.8, 0.9, 0.7, 0.85)$. We compare the LAES Algorithm with $\eta\in\{0,10,50,100,200\}$ with the UCB algorithm that makes the decision only based on the UCB estimates. Note that the LAES with $\eta=0$ coincides with the age-based scheduler. We run $500$ experiments, each of which has simulated $3\times10^4$ time slots. 

Fig. \ref{fig:sim:nonfading} shows the performance of UCB algorithm and LAES Algorithm with different $\eta$ values in the fully-connected non-fading network. We can observe from Fig. \ref{fig:sim:nonfading:regret} that UCB algorithm outperforms the LAES Algorithm with all $\eta$ values in terms of cumulative regret performance. The larger the $\eta$, the smaller the cumulative regret. This is because the larger $\eta$ puts more weight on the UCB estimates and thus the LAES Algorithm with an extremely large $\eta$ value should have similar regret as the UCB algorithm. Indeed, we can see from Fig. \ref{fig:sim:nonfading:ratio} that the packet successful delivery ratio of the best link (i.e., link $1$) increases as $\eta$ increases.  

However, the cumulative regret performance improvement is at the cost of increasing running average total age, as shown in Fig. \ref{fig:sim:nonfading:age} that shows the running average of total age over time. Indeed, we can observe from Fig. \ref{fig:sim:nonfading:age} that under the LAES Algorithm, the age becomes larger as $\eta$ increases. Nevertheless, it is worth pointing out that the age keeps increasing over time under the UCB algorithm while it is always bounded under the LAES Algorithm with all fixed $\eta$ values. We can observe similar phenomena in a relatively complicated network with $10$ links, as shown in Fig. \ref{fig:sim:fading}, despite the derived upper bound on the average total average (cf. \eqref{eqn:age:fading:ub}) in each time slot is independent of the parameter $\eta$. This is because such an upper bound is equal to $4.6\times10^7$ (much larger than $100$).


\section{Conclusion}
\label{sec:conclusion}

In this paper, we considered the problem of scheduling packets from multiple sensing sources to a central controller over a wireless network with the goal of minimizing cumulative regret over time while guaranteeing desired AoI performance. We developed a parameterized maximum-weight type scheduling policy that combines both the AoI metrics and UCB estimates in its weight measure with parameter $\eta$. We derived an upper bound on the running average total age, which linearly increases with the parameter $\eta$. We also derived an upper bound on the cumulative regret under our proposed algorithm. These derived upper bounds reveal a tradeoff: the improvement of the cumulative regret is at the cost of increasing running average total age. Simulation results were provided to confirm such a tradeoff and to demonstrate the superior performance of our proposed algorithm over the UCB algorithm and the age-based algorithm. 


\appendices
\section{Proof of Proposition \ref{prop:age}}
\label{App:proof:age}
Select the Lyapunov function 
\begin{align}
V(t) \triangleq \sum_{n=1}^{N}Z_n(t).
\end{align}
Then, under the LAES Algorithm, we have 
\begin{align}
\label{eqn:prop:age:Lyapunov}
V(t+1) =& \sum_{n=1}^{N}Z_n(t+1)\nonumber\\
\stackrel{(a)}{=}&\sum_{n=1}^{N}\left((Z_n(t)+1)(1-C_n(t)\wh{S}_n(t))+C_n(t)\wh{S}_n(t)\right)\nonumber\displaybreak[4]\\
=&\sum_{n=1}^{N}Z_n(t)-\sum_{n=1}^{N}Z_n(t)C_n(t)\wh{S}_n(t)+N\nonumber\\
\stackrel{(b)}{=}&V(t)-\sum_{n=1}^{N}Z_n(t)C_n(t)\wh{S}_n(t)+N,
\end{align}
where step $(a)$ uses the dynamics of the age (cf. \eqref{eqn:age:dynamics}) and $(b)$ follows from the definition of the Lyapunov function $V(t)$. Let $\mb{Z}(t)\triangleq(Z_n(t))_{n=1}^{N}$. Then, we have 
\begin{align}
\label{eqn:prop:age:drift}
&\bE\left[V(t+1) - V(t)\middle|\mb{Z}(t)\right]\nonumber\\
=&-\bE\left[\sum_{n=1}^{N}Z_n(t)C_n(t)\wh{S}_n(t)\middle|\mb{Z}(t)\right]+N.
\end{align}

Given the age vector $\mb{Z}(t)$ and channel state $\mb{C}(t)\triangleq(C_n(t))_{n=1}^{N}$, according to the definition of the LAES Algorithm, we have 
\begin{align}
\label{eqn:prop:age:LAES}
&\sum_{n=1}^{N}\left(Z_n(t)+\eta w_n(t)\right)C_n(t)\wh{S}_n(t)\nonumber\\
\geq&\left(Z_{n^*(t)}(t)+\eta w_{n^*(t)}\right)C_{n^*(t)}(t)\nonumber\\
\geq&Z_{n^*(t)}(t)C_{n^*(t)}(t),
\end{align}
where the first step is true for $n^*(t)\in\argmax_{n}Z_n(t)$. This implies that 
\begin{align}
\label{eqn:prop:age:LAES:1}
&\bE\left[\sum_{n=1}^{N}Z_n(t)C_n(t)\wh{S}_n(t)\middle|\mb{Z}(t)\right]\nonumber\\
\stackrel{(a)}{\geq}&\bE\left[Z_{n^*(t)}(t)C_{n^*(t)}(t)-
\eta\sum_{n=1}^{N}w_n(t)C_n(t)\wh{S}_n(t)\middle|\mb{Z}(t)\right]\nonumber\\
\stackrel{(b)}{\geq}&p_{\min}Z_{n^*(t)}(t)-\eta N,
\end{align}
where step $(a)$ uses \eqref{eqn:prop:age:LAES}, and $(b)$ uses the fact that $w_n(t)\leq 1$ $C_n(t)\leq1$ and $\wh{S}_n(t)\leq1$, $\forall n, t\geq0$ and $p_{\min}\triangleq\min_{n}p_n>0$. 

By substituting \eqref{eqn:prop:age:LAES:1} into \eqref{eqn:prop:age:drift}, we have 
\begin{align}
\label{eqn:prop:age:drift:1}
&\bE\left[V(t+1) - V(t)\middle|\mb{Z}(t)\right]\nonumber\\
\stackrel{(a)}{\leq}& -p_{\min}Z_{\max}(t)+(\eta+1)N\nonumber\\
\stackrel{(b)}{\leq}&-\frac{p_{\min}}{N}\sum_{n=1}^{N}Z_n(t)+(\eta+1)N
\end{align}
where step $(a)$ is true for $Z_{\max}(t)\triangleq\max_{n}Z_n(t)=Z_{n^*(t)}(t)$ and $(b)$ follows from the fact that $Z_{\max}(t)\geq\frac{1}{N}\sum_{n=1}^{N}Z_n(t)$.

Taking the expectation on both sides of \eqref{eqn:prop:age:drift:1}, we have 
\begin{align*}
\bE\left[V(t+1) - V(t)\right]\leq-\frac{p_{\min}}{N}\sum_{n=1}^{N}\bE\left[Z_n(t)\right]+(\eta+1)N.
\end{align*}

Summing the above inequality over time $t=0,1,\ldots,T-1$, we have 
\begin{align*}
\bE\left[V(T)-V(0)\right]\leq-\frac{p_{\min}}{N}\sum_{t=0}^{T-1}\sum_{n=1}^{N}\bE\left[Z_n(t)\right]+(\eta+1)NT,
\end{align*}
which implies  
\begin{align}
\frac{1}{T}\sum_{t=0}^{T-1}\sum_{n=1}^{N}\bE\left[Z_n(t)\right]\leq \frac{(\eta+1)N^2}{p_{\min}}.
\end{align}
Here, we use the fact that $V(0)=0$ and $V(T)\geq0$.

\section{Proof of Proposition \ref{prop:regret}}
\label{App:proof:regret}

We rewrite the regret of the LAES Algorithm as
\begin{align}
\text{Reg}(T)\triangleq& \sum_{t=0}^{T-1}\sum_{n=1}^{N}\left(\bE\left[\mu_nC_n(t)S_n^*(t)\right] -\bE\left[\mu_nC_n(t)\wh{S}_n(t)\right]\right)\nonumber\\
=&\sum_{t=0}^{T-1}\Delta R(t),
\end{align}
where $\Delta R(t)\triangleq \sum_{n=1}^{N}\bE\left[\mu_nC_n(t)S_n^*(t)-\mu_nC_n(t)\wh{S}_n(t)\right]$. 

We add the term $\eta\Delta R(t)$ on both sides of the drift of Lyapunov function $V(t)$ (cf. \eqref{eqn:prop:age:drift}) and obtain
\begin{align}
\label{eqn:prop:reg:drift}
&\bE\left[V(t+1)-V(t)\right] + \eta\Delta R(t) \nonumber\\
=&- \sum_{n=1}^{N}\bE\left[Z_n(t)C_n(t)\wh{S}_n(t)\right] + N + \eta\sum_{n=1}^{N}\bE\left[\mu_nC_n(t)S_n^*\right] \nonumber\\
&\qquad- \eta\sum_{n=1}^{N}\bE\left[\mu_nC_n(t)\wh{S}_n(t)\right]\nonumber\\
=&N+\sum_{n=1}^{N}\bE\left[\left(Z_n(t)+\eta\mu_n\right)C_n(t)\left(S_n^*-\wh{S}_n(t)\right)\right]\nonumber\\
&\qquad-\sum_{n=1}^{N}\bE\left[Z_n(t)C_n(t)S_n^*\right]\nonumber\\
\leq&N+\sum_{n=1}^{N}\bE\left[\left(Z_n(t)+\eta\mu_n\right)C_n(t)\left(S_n^*-\wh{S}_n(t)\right)\right],
\end{align}
where the last step is true since $\sum_{n=1}^{N}\bE\left[Z_n(t)C_n(t)S_n^*\right]\geq0$.

Summing \eqref{eqn:prop:reg:drift} over $t=0,1,2,\ldots,T-1$, we have 
\begin{align*}
&\sum_{t=0}^{T-1}\bE\left[V(t+1)-V(t)\right]+\eta\sum_{t=0}^{T-1}\Delta R(t)\nonumber\\
\leq&NT+\sum_{t=0}^{T-1}\sum_{n=1}^{N}\bE\left[\left(Z_n(t)+\eta\mu_n\right)C_n(t)\left(S_n^*-\wh{S}_n(t)\right)\right],
\end{align*}
which implies 
\begin{align}
\label{eqn:prop:reg:main}
&\text{Reg}(T)\triangleq\sum_{t=0}^{T-1}\Delta R(t)\leq\frac{NT}{\eta}\nonumber\\
&+\frac{1}{\eta}\sum_{t=0}^{T-1}\sum_{n=1}^{N}\bE\left[\left(Z_n(t)+\eta\mu_n\right)C_n(t)\left(S_n^*-\wh{S}_n(t)\right)\right]
\end{align}
Here, we use the fact that $V(0)=0$, $V(T)\geq0$, and the definition of $\text{Reg}(T)$.

Next, we focus on the term $$\sum_{n=1}^{N}\left(Z_n(t)+\eta\mu_n\right)C_n(t)\left(S_n^*-\wh{S}_n(t)\right).$$ Then, we have
\begin{align}
\label{eqn:prop:reg:alg}
&\sum_{n=1}^{N}\left(Z_n(t)+\eta\mu_n\right)C_n(t)\left(S_n^*-\wh{S}_n(t)\right)\nonumber\\
\stackrel{(a)}{\leq}&\sum_{n=1}^{N}\left(Z_n(t)+\eta\mu_n\right)C_n(t)\wt{S}_n(t)\nonumber\\
&\qquad\qquad-\sum_{n=1}^{N}\left(Z_n(t)+\eta\mu_n\right)C_n(t)\wh{S}_n(t)\nonumber\\
\stackrel{(b)}{\leq}&\sum_{n=1}^{N}\left(Z_n(t)+\eta\mu_n\right)C_n(t)\wt{S}_n(t)\nonumber\\
&\qquad\qquad-\sum_{n=1}^{N}\left(Z_n(t)+\eta\mu_n\right)C_n(t)\wh{S}_n(t)\nonumber\\
&+\sum_{n=1}^{N}\left(Z_n(t)+\eta w_n(t)\right)C_n(t)\wh{S}_n(t)\nonumber\\
&\qquad\qquad-\sum_{n=1}^{N}\left(Z_n(t)+\eta w_n(t)\right)C_n(t)\wt{S}_n(t)\nonumber\\
=&\eta\sum_{n=1}^{N}(w_n(t)-\mu_n)C_n(t)\wh{S}_n(t)\nonumber\\
&\qquad\qquad +\eta\sum_{n=1}^{N}(\mu_n-w_n(t))C_n(t)\wt{S}_n(t),
\end{align}
where step $(a)$ is true for $$\wt{\mb{S}}(t)\triangleq(\wt{S}_n(t))_{n=1}^{N}\in\argmax_{\mb{S}\in\mc{S}}\sum_{n=1}^{N}\left(Z_n(t)+\eta\mu_n\right)C_n(t)S_n,$$
and $(b)$ uses the definition of $\wh{\mb{S}}(t)$.

By substituting \eqref{eqn:prop:reg:alg} into \eqref{eqn:prop:reg:main}, we have 
\begin{align}
\label{eqn:prop:reg:final}
&\text{Reg}(T)\leq\frac{NT}{\eta}+\underbrace{\sum_{t=0}^{T-1}\sum_{n=1}^{N}\bE\left[\left(w_n(t)-\mu_n\right)C_n(t)\wh{S}_n(t)\right]}_{\triangleq G_1(T)} \nonumber\\
&\qquad\qquad+ \underbrace{\sum_{t=0}^{T-1}\sum_{n=1}^{N}\bE\left[\left(\mu_n-w_n(t)\right)C_n(t)\wt{S}_n(t)\right]}_{\triangleq G_2(T)}.
\end{align}


Next, we focus on $G_1(T)$ and $G_2(T)$, respectively. Let $t_{n,\tau}$ denote the time slot at which link $n$ successfully received a packet, i.e., $C_n(t_{n,\tau})\wh{S}_n(t_{n,\tau})=1$ and $C_n(t_{n,\tau})\wh{S}_n(t_{n,\tau})=0$ if $t\neq t_{n,\tau}, \tau=1,2,\ldots, H_n(T)$. Therefore, we have $H_n(t_{n,\tau})=\tau-1$.

Let $G_{n,1}(T)\triangleq\sum_{t=0}^{T-1}\bE\left[\left(w_n(t)-\mu_n\right)C_n(t)\wh{S}_n(t)\right] $ and thus $G_1(T)=\sum_{n=1}^{N}G_{n,1}(T)$.

Hence, we have 
\begin{align}
\label{eqn:prop:reg:R1}
G_{n,1}(T)\stackrel{(a)}{\leq}&\sum_{t=0}^{T-1}\bE\left[(w_n(t)-\mu_n)C_n(t)\wh{S}_n(t)\id_{\mc{F}_n(t)}\right] \nonumber\\
\stackrel{(b)}{\leq}&\bE\left[\sum_{\tau=1}^{H_n(T)}(w_n(t_{n,\tau})-\mu_n)\id_{\mc{F}_n(t_{n,\tau})}\right] \nonumber\\
\stackrel{(c)}{\leq}&1+\bE\left[\sum_{\tau=2}^{H_n(T)}(w_n(t_{n,\tau})-\mu_n)\id_{\mc{F}_n(t_{n,\tau})}\right]\nonumber\\
\stackrel{(d)}{\leq}&1+\bE\left[\sum_{\tau=2}^{H_n(T)}(w_n(t_{n,\tau})-\mu_n)\id_{\mc{F}_n(t_{n,\tau})\cap\mc{G}_n(t_{n,\tau})}\right]\nonumber\\
&\quad+\bE\left[\sum_{\tau=2}^{H_n(T)}\id_{\ol{\mc{G}}_n(t_{n,\tau)}}\right],
\end{align}
where step $(a)$ is true for $\mc{F}_n(t)\triangleq\{w_n(t)\geq\mu_n\}$ and $\id_{\{\cdot\}}$ being an indicator function; $(b)$ uses the definition of $t_{n,\tau}$, and the fact that $C_n(t)\leq1$ and $\wh{S}_n(t)\leq1,\forall t\geq0$; $(c)$ follows from the fact that $w_n(t)\leq1, \forall t\geq0$; $(d)$ is true for 
$$\mc{G}_n(t)\triangleq\left\{\ol{\mu}_n(t)-\mu_n\leq\sqrt{\frac{3\log t}{2H_n(t)}}\right\},$$
and $\ol{\mc{G}}_n(t)$ being the complement of the event $\mc{G}_n(t)$.

Next, we consider the second term on the right hand side (RHS) of \eqref{eqn:prop:reg:R1}. 
\begin{align}
\label{eqn:prop:reg:R1:second}
&\bE\left[\sum_{\tau=2}^{H_n(T)}(w_n(t_{n,\tau})-\mu_n)\id_{\mc{F}_n(t_{n,\tau})\cap\mc{G}_n(t_{n,\tau})}\right] \nonumber\\
\stackrel{(a)}{\leq}&\bE\left[\sum_{\tau=2}^{H_n(T)}2\sqrt{\frac{3\log t_{n,\tau}}{2H_n(t_{n,\tau})}}\right] \nonumber\\
\stackrel{(b)}{\leq}&\sqrt{6\log T}\bE\left[\sum_{\tau=2}^{H_n(T)}\frac{1}{\sqrt{\tau-1}}\right]\nonumber\\
\leq&\sqrt{6\log T}\left(1 + \int_{1}^{H_n(T)}\frac{1}{\sqrt{x}}dx\right)\nonumber\\
\leq&2\sqrt{6\log T}\bE\left[\sqrt{H_n(T)}\right],
\end{align}
where step $(a)$ uses the definition of $w_n(t)$ and $\mc{G}_n(t)$, and $(b)$ follows from the fact that $t_{n,\tau}\leq T$ and the definition of $t_{n,\tau}$. With regard to the third term on the RHS of \eqref{eqn:prop:reg:R1}, we have 
\begin{align*}
&\bE\left[\id_{\ol{\mc{G}}_n(t_{n,\tau})}\right]=\Pr\{\ol{\mc{G}}_n(t_{n,\tau})\}\nonumber\\
\stackrel{(a)}{\leq}&\Pr\left\{\bigcup_{m=\tau-1}^{T-1}\left\{\ol{\mu}_n(m)-\mu_n>\sqrt{\frac{3\log m}{2(\tau-1)}}\right\}\right\}\nonumber\\
\leq&\Pr\left\{\bigcup_{m=\tau-1}^{T-1}\left\{\ol{\mu}_n(m)-\mu_n>\sqrt{\frac{3\log m}{2m}}\right\}\right\}\nonumber\displaybreak[4]\\
\stackrel{(b)}{\leq}&\sum_{m=\tau-1}^{T-1}\Pr\left\{\ol{\mu}_n(m)-\mu_n>\sqrt{\frac{3\log m}{2m}}\right\} \nonumber\\
\stackrel{(c)}{\leq}&\sum_{m=\tau-1}^{T-1}\frac{1}{m^3}\leq\frac{1}{(\tau-1)^3}+\int_{\tau-1}^{\infty}\frac{1}{x^3}dx\stackrel{(d)}{\leq}\frac{3}{2(\tau-1)^2},
\end{align*}
where step $(a)$ follows from the fact that 
$$\ol{\mc{G}}_n(t_{n,\tau})\subset\bigcup_{m=\tau-1}^{T-1}\left\{\ol{\mu}_n(m)-\mu_n>\sqrt{\frac{3\log m}{2(\tau-1)}}\right\};$$
$(b)$ uses the union bound; $(c)$ follows from the Chernoff-Hoeffding Bound (see, e.g., \cite[Fact 1]{auer2002finite}), i.e., for $X_1,X_2,\ldots,X_n$ be i.i.d. random variables with common range $[0,1]$ and mean $\mu$, then for any $a\geq0$, we have 
\begin{align}
\label{eqn:prop:reg:chernoff}
\Pr\left\{\frac{1}{n}\sum_{i=1}^{n}X_i\geq \mu+a\right\}\leq e^{-2na^2},
\end{align}
$(d)$ is true for $\tau\geq2$.

Hence, the third term on the RHS of \eqref{eqn:prop:reg:R1} can be bounded as follows.
\begin{align}
\label{eqn:prop:reg:R1:third}
\bE\left[\sum_{\tau=2}^{H_n(T)}\id_{\ol{\mc{G}}_n(t_{n,\tau})}\right]\leq&\bE\left[\sum_{\tau=2}^{H_n(T)}\frac{3}{2(\tau-1)^2}\right]\nonumber\\
\leq&\sum_{\tau=1}^{\infty}\frac{3}{2\tau^2}=\frac{\pi^2}{4},
\end{align}
where the last step use the fact that $\sum_{n=1}^{\infty}1/n^2=\pi^2/6$. By substituting \eqref{eqn:prop:reg:R1:second} and \eqref{eqn:prop:reg:R1:third} into \eqref{eqn:prop:reg:R1} and using the definition of $G_1(T)$, we have 
\begin{align}
\label{eqn:prop:reg:R1:final}
&G_1(T)\leq N\left(1+\frac{\pi^2}{4}\right) + 2\sqrt{6\log T}\sum_{n=1}^{N}\bE\left[\sqrt{H_n(T)}\right]\nonumber\\
\stackrel{(a)}{\leq}&N\left(1+\frac{\pi^2}{4}\right) + 2N\sqrt{6\log T}\bE\left[\sqrt{\frac{1}{N}\sum_{n=1}^{N}H_n(T)}\right] \nonumber\\
\stackrel{(b)}{\leq}&N\left(1+\frac{\pi^2}{4}\right) + 2\sqrt{6N|\mb{S}|_{\max}T\log T},
\end{align}
where step $(a)$ uses the Jensen's inequality, and $(b)$ is true since $\sum_{n=1}^{N}H_n(T)\leq T|\mb{S}|_{\max}$ and $|\mb{S}|_{\max}$ is the maximum number of links that can be scheduled in each time slot.

Next, we consider the term $G_2(T)$. First, we note that  
\begin{align}
G_2(T)\leq\sum_{t=0}^{T-1}\sum_{n=1}^{N}\bE\left[(\mu_n-w_n(t))\wt{S}_n(t)\id_{\ol{\mc{F}}_n(t)}\right],
\end{align}
where we recall that $\mc{F}_n(t)\triangleq\{w_n(t) \geq \mu_n\}$. Note that for $t\leq t_{n,1}$, we have $w_n(t)=1$ and thus $\mc{F}_n(t)$ happens. Therefore, we have 
\begin{align}
\label{eqn:prop:reg:R2:final}
&G_2(T)\leq\sum_{n=1}^{N}\bE\left[\sum_{t=t_{n,1}+1}^{T-1}\left(\mu_n-w_n(t)\right)\wt{S}_n(t)\id_{\ol{\mc{F}}_n(t)}\right]\nonumber\\
\stackrel{(a)}{\leq}&\sum_{n=1}^{N}\bE\left[\sum_{t=t_{n,1}+1}^{T-1}\Pr\left\{\ol{\mu}_n(t)-\mu_n\leq-\sqrt{\frac{3\log t}{2H_n(t-1)}}\right\}\right]\nonumber\\
\leq&\sum_{n=1}^{N}\sum_{\tau=1}^{T-1}\sum_{m=1}^{\tau}\Pr\left\{\frac{1}{m}\sum_{i=1}^{m}X(i)-\mu_n\leq-\sqrt{\frac{3\log \tau}{2m}}\right\}\nonumber\\
\stackrel{(b)}{\leq}&\sum_{n=1}^{N}\sum_{\tau=1}^{T-1}\sum_{m=1}^{\tau}\frac{1}{\tau^3}=\sum_{n=1}^{N}\sum_{\tau=1}^{T-1}\frac{1}{\tau^2}\stackrel{(c)}{\leq}\frac{N\pi^2}{6},
\end{align}
where step $(a)$ follows from the fact that $\mu_n\leq 1$ and $\wt{S}_n(t)\leq 1$ as well as the definition of $\ol{\mc{F}}_n(t)$; $(b)$ again uses the Chernoff-Hoeffding Bound (cf. \eqref{eqn:prop:reg:chernoff}); $(c)$ is true since $\sum_{\tau=1}^{T-1}1/\tau^2\leq\sum_{\tau=1}^{\infty}1/\tau^2=\pi^2/6$. 

Hence, by substituting \eqref{eqn:prop:reg:R1:final} and \eqref{eqn:prop:reg:R2:final} into \eqref{eqn:prop:reg:final}, we have the desired result.




\bibliographystyle{IEEEtran}
\bibliography{refs}

\end{document}